\documentclass[twocolumn,twoside,slac_two]{revtex4}
\usepackage{graphicx}
\usepackage{fancyhdr}
\usepackage{amsmath}
\usepackage{amsfonts}
\usepackage{amssymb}
\newcommand{\E}[1]{\ensuremath{\langle #1 \rangle}}
\newcommand{\sigl}{\ensuremath{\sigma_{\!\mathcal{L}}}}

\begin{document}

\title{Techniques for targeted Fermi-GBM follow-up of gravitational-wave events}


\author{L. Blackburn}
\altaffiliation{NASA Postdoctoral Program Fellow}
\affiliation{NASA/Goddard Space Flight Center, Greenbelt, MD, 20771, USA}
\author{M. S. Briggs}
\affiliation{University of Alabama in Huntsville, Huntsville, AL, 35805, USA}
\author{J. Camp}
\affiliation{NASA/Goddard Space Flight Center, Greenbelt, MD, 20771, USA}
\author{V. Connaughton}
\affiliation{University of Alabama in Huntsville, Huntsville, AL, 35805, USA}
\author{N. Christensen}
\affiliation{Carleton College, Northfield, MN, 55057, USA}
\author{P. Jenke}
\affiliation{University of Alabama in Huntsville, Huntsville, AL, 35805, USA}
\author{J. Veitch}
\affiliation{NIKHEF, Science Park 105, Amsterdam 1098XG, The Netherlands}

\begin{abstract}
    The Advanced LIGO and Advanced Virgo ground-based gravitational-wave (GW)
    detectors are projected to come online 2015--2016, reaching a final
    sensitivity sufficient to observe dozens of binary neutron star mergers per
    year by 2018. We present a fully-automated, targeted search strategy for
    prompt gamma-ray counterparts in offline Fermi-GBM data. The multi-detector
    method makes use of a detailed model response of the instrument, and
    benefits from time and sky location information derived from the
    gravitational-wave signal.
\end{abstract}

\maketitle

\thispagestyle{fancy}

% References should be done using the \citep, \ref, and \label commands

\section{Introduction}

Compact binary coalescence (CBC), such as the merger of two neutron stars (NS)
or black holes (BH), remains the most highly anticipated gravitational-wave
signal for ground-based gravitational-wave detectors. The second-generation
Advanced LIGO (4km-baseline interferometers in Hanford, WA and Livingston, LA)
\citep{Harry:2010zz} and Advanced Virgo (3km interferometer in Cascina, Italy)
\citep{Accadia:2009zz} detectors are projected to come online 2015--2016, and
reach a final sensitivity sufficient to observe dozens of NS-NS mergers per
year by 2018. Together with the 600m GEO-HF detector \citep{Willke:2006uw}, they
will form a word-wide network of gravitational-wave interferometric detectors.
The network will be joined by the Japanese 3km cryogenic KAGRA detector
\citep{Somiya:2011np} and a proposed third LIGO-India detector around 2018--2020,
increasing overall sensitivity and sky-coverage, while also improving sky
localization and waveform reconstruction \citep{Schutz:2011tw}.

Gravitational-waves have yet to be directly observed. Our confidence in the
existence of compact binary coalescence events comes primarily from the
discovery of a small number of galactic pulsars which appear to be in close
binary systems with another neutron star. The first and most famous of these
systems contains the Hulse-Taylor pulsar PSR B1316+16. It has demonstrated over
many decades an orbital decay consistent with loss of energy to gravitational
waves~\citep{Weisberg:2010zz}.  The number and inferred lifetime of these
systems can be used to obtain an estimate of about one NS-NS merger event per
Mpc$^3$ per million years~\citep{Abadie:2010cf}, with up to two orders of
magnitude uncertainty in rate due largely to our limited knowledge of the
pulsar luminosity function and limited statistics. The merger rate translates
into an estimate of $\sim$0.02 detectable NS-NS merger events per year for the
initial LIGO-Virgo detector network, which operated between 2005--2010, and
$\sim$40 per year for the advanced detectors once they reach design
sensitivity. Although their gravitational radiation is stronger, the merger
rates of NS-BH binaries is more uncertain as we have not observed any NS-BH
binary systems, and have generally poor knowledge of the black hole mass
distribution.

Gamma-ray bursts (GRB) are flashes of gamma rays observed approximately once
per day. Their isotropic distribution in the sky was the first evidence of an
extra-galactic origin, and indicated that they were extremely energetic events.
The duration of prompt gamma-ray emission shows a bi-modal distribution which
naturally groups GRBs into two categories \citep{Kouveliotou:1993yx}. Most long
GRBs emit their prompt radiation over timescales $\gtrsim$2 seconds, and as
much as hundreds of seconds. They have been associated with young stellar
populations and the collapse of rapidly rotating massive stars
\citep{Hjorth:2011zx}. Short GRBs (sGRB), with prompt emission typically less
than 1s and a generally harder spectrum, are found in both old and new stellar
populations. Mergers of two neutron stars, or of neutron star/black hole
systems, are thought thought to be a major contribution to the sGRB population
\citep{Nakar:2007yr}. It is this favored progenitor model which makes short GRBs
and associated afterglow emission a promising counterpart to gravitational-wave
observations.

Since 2005, the Swift satellite has revolutionized our understanding of short
GRBs by the rapid observation of x-ray afterglows, providing the first
localization, host identification, and red-shift
information~\citep{Gehrels:2009qy}. The beaming angle for short GRBs is highly
uncertain, although limited observations of jet breaks in some afterglows imply
half-opening angles of $\theta_j \sim$ 3--14 degrees
\citep{Liang:2007rn,Fong:2012aq}. The absence of an observable jet break sets a
lower limit on the opening angle which is generally weak (due to limits in
sensitivity), though in the case of GRB 050724A, late-time Chandra observations
were able to constrain $\theta_j \gtrsim 25^\circ$~ \citep{Grupe:2006uc}.

The observed spatial density of sGRB's and limits on beaming angle result in a
NS-NS merger event rate roughly consistent with that derived from galactic
binary pulsar measurements. Although the beaming factor of $\sim\theta_j^2/2$
means we believe most merger events seen by the advanced GW detectors will not
be accompanied by a standard gamma-ray burst, this is somewhat compensated by
the fact that the ones that are beamed toward us have stronger
gravitational-wave emission. Current estimates for coincident GW-sGRB
observation for advanced LIGO-Virgo are a few per year assuming a NS-NS
progenitor model \citep{Metzger:2011bv,Coward:2012gn}.  The rate increases by a
factor of 8 if all observed short GRBs are instead due to NS-BH (10 $M_\odot$)
mergers which are detectable in gravitational-waves to about twice the
distance.

In addition to the jet-driven burst and afterglow, other EM emission associated
with a compact merger can be a promising channel for GW-EM coincidence,
particularly if the EM radiation is less-beamed or even isotropic. A few short
GRBs ($\sim$10\%) have shown clear evidence of high-energy flares which precede
the primary burst by 1--10 seconds, and possibly up to
100s~\citep{Troja:2010zm}. The precursors can be interpreted as evidence of some
activity during or before merger, such as the resonant shattering of NS crusts
\citep{Tsang:2011ad}, which could radiate isotropically. Thus it will be
interesting to search for weak non-standard EM emission accompanying all nearby
NS-NS mergers seen in GWs, while we expect only a small fraction to be
oriented in our line-of-sight for a standard jet-driven sGRB.

The Gamma-ray Burst Monitor (GBM) \citep{Meegan:2009qu} aboard the Fermi
spacecraft measures photon rates from 8 keV--40 MeV. The instrument consists of
12 semi-directional NaI scintillation detectors and 2 BGO scintillation
detectors which cover the entire sky not occluded by the Earth (about 65\%).
The lower-energy NaI detectors have an approximately $\cos{\theta}$ response
relative to angle of incidence, and relative rates across detectors are used to
reconstruct the source location to a few degrees.  The BGO detectors are much
less directional, and are used to detect and resolve the higher energy spectrum
above $\sim$200 MeV.

GBM produces on-board triggers for gamma-ray burst events by looking for
multi-detector rate excess over background across various energy bands and
timescales. In the case of a trigger, individual photon information is sent to
the ground and the event is publicly reported. Those events which have been
confirmed as GRBs have already been studied in coincidence with LIGO-Virgo data
\citep{Abbott:2009kk,Abadie:2010uf,Briggs:2012ce}. So far, no gravitational-wave
counterparts to triggered GRBs have been identified, which was not unexpected
given the limited reach of the first generation instruments.

In addition to the triggered events, survey data is available which records
binned photon counts over all time. In this proposed offline analysis of short
transients, we consider the CTIME daily data, which contains counts binned at
0.256s over 8 energy channels for each detector. A new GBM data product
(continuous TTE) was implemented in late 2012. It provides continuous data on
individual photons with 2$\mu$s and 128 energy channel resolution, which will
further enhance offline sensitivity to particularly short bursts.

This paper discusses the possibility of using these offline GBM data products
to follow-up gravitational-wave candidates in the advanced LIGO-Virgo era.  In
section \ref{sec:gw} we describe the characteristics of a trigger provided by
gravitational-wave data.  Sections \ref{sec:gbm} and \ref{sec:followup} develop
a likelihood-ratio based procedure for analyzing available GBM offline data
about the time and sky location provided by the gravitational waves. Finally,
in section \ref{sec:test} we demonstrate the performance of the search
algorithm on background times and a sample of known sGRB's.

\section{Gravitational-wave trigger}
\label{sec:gw}

Searches for NS-NS and NS-BH coalescence in gravitational-wave data typically
use matched filtering of the model waveform
\citep{Colaboration:2011np,LIGO:2012aa}. The gravitational waves from CBC are
characterized by a chirp of monotonically increasing frequency and amplitude as
the binary inspirals under approximately adiabatic orbital decay. During this
progression, kilometer-scale interferometers are sensitive to the inspiral phase
of a stellar-mass coalescence until just prior to merger, but before any tidal
effects on a NS become important.

\begin{figure}
\begin{center}
        \includegraphics[width=2.5in]{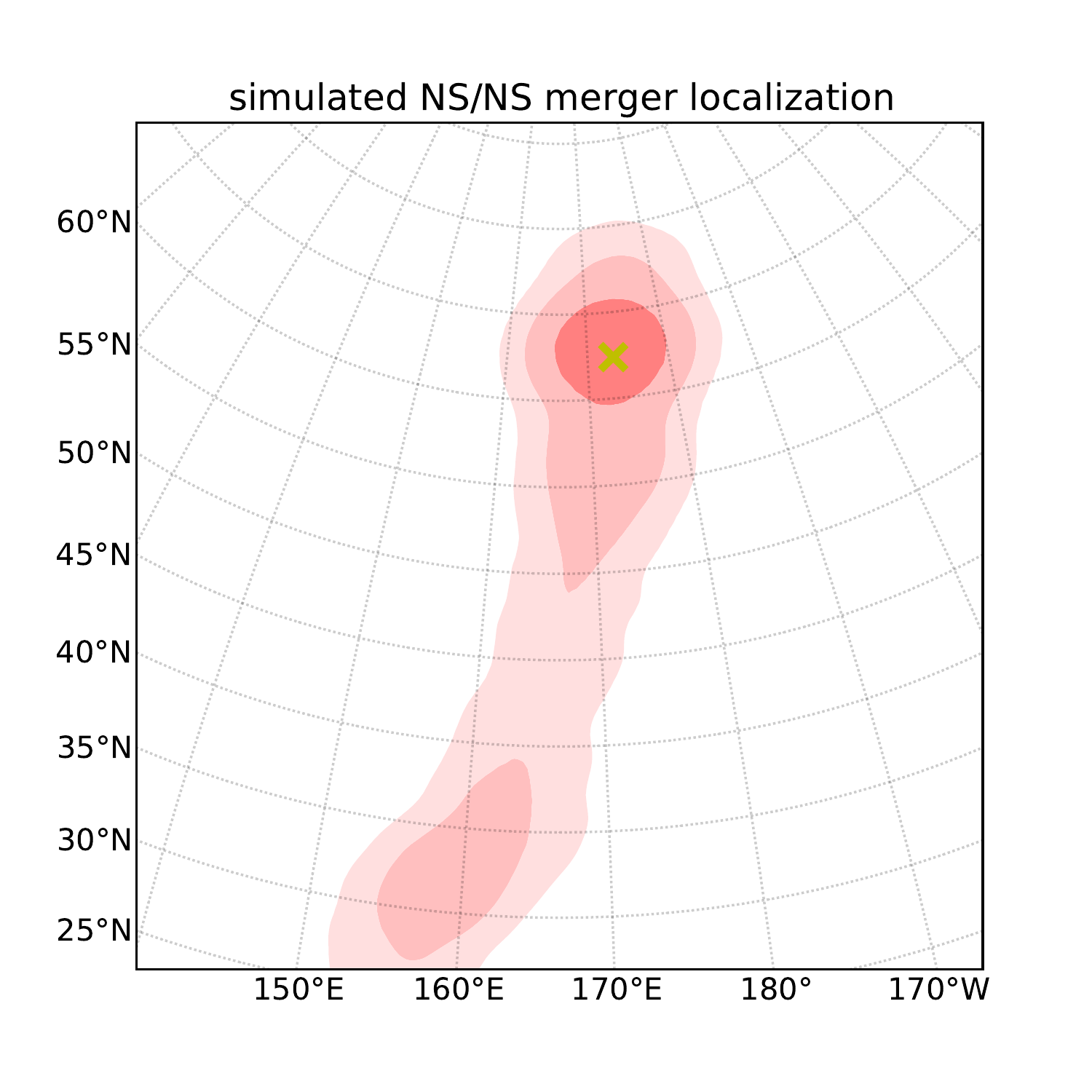} \caption{Example
            sky localization of simulated NS-NS merger seen in modeled
            gravitational-wave data with noise similar in spectrum to that
            expected from the advanced LIGO-Virgo GW network. The
            signal-to-noise of the GW signal in each detector is 8.7 (H1) 7.2
            (L1) and 3.1 (V1), and the sky location is determined through
            Bayesian inference using a nested sampling technique
            \citep{Veitch:2012df} applied to data from all detectors. The
            elongated shape beyond 1$\sigma$ is due to a small degenerate
            timing uncertainty in V1.}
        \label{fig:localization}
\end{center}
\end{figure}

The waveform in this regime is well-modeled using post-Newtonian techniques,
and can be considered a standard candle encoding orbital parameters of the
system. Mass, spin, and coalescence time are accessible with the GW projection
onto a single detector, while distance, inclination, and sky location
degeneracies can be disentangled to varying degrees of success using coherent
data from multiple detectors, for example by using Bayesian inference
\citep{Veitch:2012df}. A typical gravitational-wave detection in the early
advanced detector era may have moderate signal-to-noise $\sim$8 in two or more
detectors, estimated merger time to within milliseconds, and sky localization
to $\sim$100 square degrees depending on detector network configuration
\citep{Fairhurst:2010is}.  Figure \ref{fig:localization} shows a reconstructed
3-detector localization for a simulated signal near detection threshold.

\section{Coherent analysis of GBM data}
\label{sec:gbm}

In this section, we develop a procedure to coherently search GBM detector data
for modeled events. The basic idea is that by processing multiple detector data
coherently, we can obtain a greater sensitivity than when considering one
detector at a time. Greater computational resources available offline (vs.
on-board) also allow for more careful background estimation to be done.  For
this analysis, we can relax to some extent the strict 2-detector coincidence
requirement used to veto spurious events on-board as the gravitational-wave
trigger means much less time and sky area is considered.

\subsection{GBM background estimation}

\begin{figure}
    \begin{center}
        \includegraphics[width=3in]{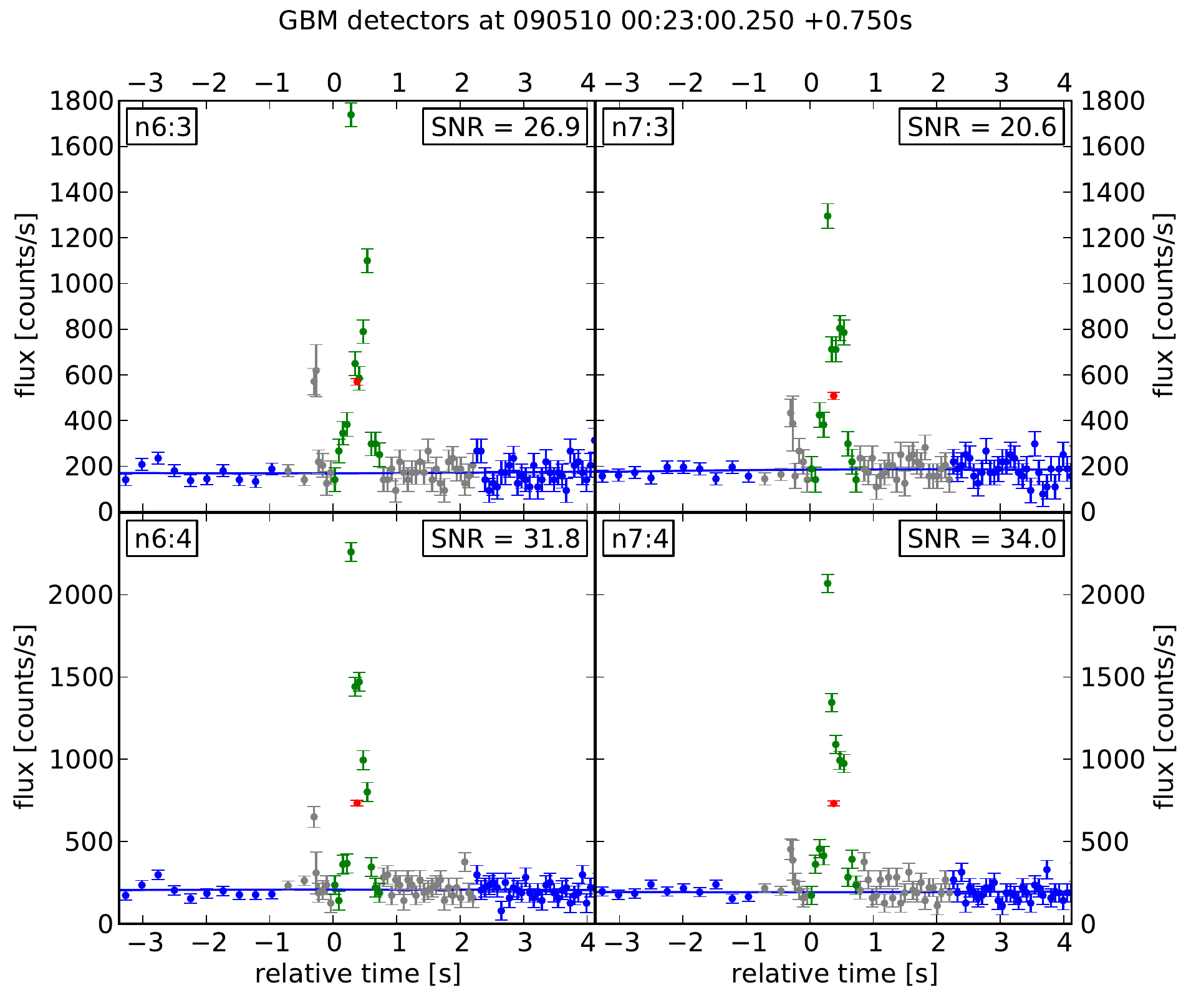} \caption{Detector and
            channel-dependent polynomial background fitting applied to GRB
            090510A. Four of the 8 $\times$ 14 channel and detector
            combinations are shown. The green represents the highest likelihood
            foreground interval, and the background contribution is estimated
            using a polynomial fit to the region (cropped) shown in blue. Red
            dots mark the average flux in the foreground interval.}
        \label{fig:bgfit}
    \end{center}
\end{figure}

Each detector is subject to a substantial time-varying background from bright
high-energy sources that come in and out of the wide field of view, as well as
location-dependent particle and Earth atmospheric effects. This background must
be estimated and subtracted out to look for any prompt excess.  Methods in use
include the local averaging of previous data done on-board, smooth spline-fits
with a high-frequency cutoff, direct tracking and modeling of the dominant
sources, and averaging rates from previous orbits with similar
orientations~\citep{Finger:2009id,WilsonHodge:2012ix,Fitzpatrick:2012np}. In
this analysis where we are interested in the background estimate for a short
foreground interval [$-T/2$, $+T/2$] where $T \sim$1s, we estimate the
background using a polynomial fit to local data from [$-10T$, $+10T$] (minimum
$\pm5s$), excluding time [$-3T/2$, $+5T/2$] around the foreground interval to
avoid bias from an on-source excess.  An example of the foreground and
background intervals about a strong prompt signal is shown in figure
\ref{fig:bgfit}. The polynomial degree is determined by the interval length to
account for more complicated background variability over longer intervals. It
ranges from 2 (minimum) to 1+$0.5\log_2 T$. The quality of the fit is
determined using a $\chi^2$ statistic applied to the data, re-binned at $T/4$
seconds for $T>1$s.  If the $\chi^2$ per degree-of-freedom is over 2, or if any
of the $N$ individual data points has $\chi^2 > 4\ln N$, it is assumed that the
polynomial could not adequately model the local background variability, and the
fit is redone over the smaller interval [$-5T$, $+5T$]. If the fit continues to
fail the $\chi^2$ test at the looser requirements of 3 and $6\ln N$
respectively, the background estimate is marked unreliable and not used in
further computations for that particular foreground interval.

High-energy cosmic rays striking a NaI crystal can result in long-lived
phosphorescent light emission. The detector may interpret this is a rapid
series of events, creating a short-lived jump in rates for one or multiple
channels, and severely distorting the background fit if not accounted for.
They are identified with a simple procedure that compares the counts in each
0.256s bin against the mean rate estimated from four neighboring bins. If an
excess is detected to 5$\sigma$ or more, that measurement and the immediately
neighboring bin on either side is removed from the background fit for all
channels of the affected detector. A much stricter requirement is used when
rejecting cosmic-ray events in the foreground interval. In that case, a 0.256s
measurement must have signal-to-noise over $500/\sqrt{T}$ for it to be excised,
which is much larger than expected for true GRBs.

\subsection{Likelihood-ratio statistic}

A likelihood ratio combines information about sources and noise into a single
variable. It is defined as the probability of measuring the observed data, $d$, in
the presence of a particular true signal $H_1$ (source amplitude $s > 0$)
divided by the probability of measuring the observed data in noise alone $H_0$
($s = 0$).
\begin{equation}
    \Lambda(d) = \frac{P(d | H_1)}{P(d | H_0)}
\end{equation}
When signal parameters such as light-curve, spectrum, amplitude $s$ and
sky-location $\alpha, \delta$ are unknown, one can either marginalize over the
unknown parameters, or take the maximum likelihood over the range to obtain
best-fit values.

For $i$ measurements of sufficiently binned, uncorrelated Gaussian data,
\begin{align}
    \label{eqn:lrnumerator}
    P(d_i|H_1) &= \prod_i{\frac{1}{\sqrt{2\pi}\sigma_{d_i}}
        \exp\left(-\frac{(\tilde{d_i}-r_is)^2}{2\sigma_{d_i}^2}\right)} \\
    \label{eqn:lrdenominator}
    P(d_i|H_0) &= \prod_i{\frac{1}{\sqrt{2\pi}\sigma_{n_i}}
        \exp\left(-\frac{\tilde{d_i}^2}{2\sigma_{n_i}^2}\right)}
\end{align}
where we have used $\tilde{d_i} = d_i - \E{n_i}$ to represent the
background-subtracted measurements (e.g. red dots minus blue curve in figure
\ref{fig:bgfit}), $\sigma_{n_i}$ and $\sigma_{d_i}$ for the standard deviation
of the background and expected data (background+signal), $r_i$ for the
location/spectrum-dependent instrumental response, and $s$, a single intrinsic
source amplitude scaling factor at the Earth. Maximizing the likelihood ratio
is the same as maximizing the log-likelihood ratio $\mathcal{L} =
\ln{\Lambda}$,
\begin{equation}
    \label{eqn:loglikelihood}
    \mathcal{L} = \sum_i{\left[\ln{\frac{\sigma_{n_i}}{\sigma_{d_i}}}
            +\frac{\tilde{d_i}^2}{2\sigma_{n_i}^2}
            -\frac{(\tilde{d_i}-r_is)^2}{2\sigma_{d_i}^2}\right]}
\end{equation}

When the first two terms are fixed, maximizing the log-likelihood ratio is
equivalent to minimizing the the third term which we recognize as a $\chi^2$
fit to the two free sky-location parameters in $r_i$ and the single amplitude
parameter $s$. The shape of the likelihood function over source amplitude $s$
is the product of (almost) Gaussians centered about the best-estimate of $s$ in
each measurement, and the total likelihood is scaled by the sum of the squared
signal-to-noise of all measurements.

The dependence of response factors $r_i$ on sky location is complicated, so the
likelihood ratio is calculated over a sample grid of all possible locations.
Assuming a single location, the remaining free parameter is the source
amplitude $s$. The variance in the background-subtracted detector data includes
both background and source contributions,
\begin{equation}
    \label{eqn:sigmadi}
    \sigma_{d_i}^2 = \sigma_{n_i}^2 + r_is + \sigma_{r_i}^2s^2 \quad (s \ge 0)
\end{equation}
with $\sigma_{r_i}^2$ representing Gaussian-modeled systematic uncertainty in
the instrumental response. Source terms are only included for physical $s \ge
0$ else their contribution is zero. The background contributes Poisson error,
as well as any systematic variance $\sigma_{b_i}^2$ from poor background
fitting which is also assumed to be Gaussian,
\begin{equation}
    \label{eqn:sigmani}
    \sigma_{n_i}^2 = \E{n_i} + \sigma_{b_i}^2.
\end{equation}
We find $s_\mathrm{best}$ which maximizes $\mathcal{L}$ by setting the
derivative $d\mathcal{L}/ds$ to zero. If $\sigma_{d_i} = \sigma_{n_i}$, which
happens when $r s \ll \sigma_{n}^2$, $\sqrt{2 \mathcal{L}_\mathrm{max}}$
reduces to a coherent SNR (sum data using weights $r_i/\sigma_{n_i}^2$).  If
$\sigma_{d_i}$ can be assumed constant, the solution for $s_\mathrm{best}$ is
found analytically,
\begin{equation}
    \label{eqn:esguess}
    s_\mathrm{best} \approx \frac{\sum_i{r_i \tilde{d_i} / \sigma_{d_i}^2}}{\sum_i{r_i^2 / \sigma_{d_i}^2}},
\end{equation}
which is just an appropriately inverse-noise weighted sum of the individual
estimates of $s$ from each measurement.  Although $\sigma_{d_i}$ depends on
$s$, we can make the practical approximation $\sigma_{d_i}^2 \approx
\max(\E{n_i}, d_i) + \sigma_{b_i}^2$ as an initial guess.

To find the true maximum, we solve for $d\mathcal{L}/ds = 0$ using Newton's
method, beginning with the value of the first derivative at our initial guess
$s_0$, and using the analytic second derivative to refine the measurement:
$s_\mathrm{best} \approx s_0 - (\partial\mathcal{L}/\partial s)/(\partial^2\mathcal{L}/\partial s^2)$.
This calculation is both fast and easily vectorized. One initial guess using equation
\ref{eqn:esguess} followed by a couple iterations of Newton's method generally
provides an excellent approximation for $s_\mathrm{best}$.

So far we have been calculating the probability of a signal assuming a specific
source amplitude $s$. To consider all possible source amplitudes we need to
integrate the likelihood $P(d|s)$ (equation \ref{eqn:lrnumerator}) over a prior
on $s$,
\begin{equation}
    \label{eqn:marginalization}
    P(d) = \int{P(d|s)P(s)ds}
\end{equation}
For a given set of detector data $d$, the likelihood $P(d|s)$ over $s$ is
almost the product of individual Gaussian distributions (not quite Gaussian
because $\sigma_d$ depends on $s$). The product of Gaussian distributions with
mean values $\mu_i$ and standard deviations $\sigma_i$ is itself Gaussian with
mean and variance,
\begin{equation}
    \mu_\mathrm{prod} = \frac{\sum{\mu_i/\sigma^2_i}}{\sum{1/\sigma^2_i}}, \quad
    \sigma^2_\mathrm{prod} = \frac{1}{\sum{1/\sigma^2_i}}
\end{equation}
In this case, $\mu_i = \tilde{d_i}/r_i$ and $\sigma_i = \sigma_{d_i} / r_i$.
The mean value $\mu_\mathrm{prod}$ is the same as our initial guess for $s$ at
maximum likelihood (equation \ref{eqn:esguess}), but we can assume to have a
more accurate maximum-likelihood location $s_\mathrm{best}$ from the numerical
procedure outlined above. The estimate for the variance of $\mathcal{L}$ over
$s$ is,
\begin{equation}
    \sigl^2 = \frac{1}{\sum{r_i^2/\sigma_{d_i}^2}}, \quad \sigma^2_{d_i} \text{ evaluated at } s_\mathrm{best}
    \label{eqn:sigl}
\end{equation}

For a flat prior $P(s) = 1$, we can integrate
equation~\ref{eqn:marginalization} by simply considering the area of a Gaussian
with peak value $P(d|s_\mathrm{best})$ and variance $\sigl^2$.  Other choices
for a prior can be represented by a power-law distribution $P(s) \propto
s^{-\beta}$.  A spatially homogeneous population, suitable for nearby sources,
would assume $\beta = 5/2$, while the empirical distribution for observed GRB
amplitudes follows a power-law decay closer to $\beta \approx 1.8$, reflecting
cosmological effects. If instead we are looking for signals from a particular
host galaxy at known location and distance, we want to use an intrinsic source
luminosity distribution. A convenient option is to use a scale-free prior with
fixed $\beta = 1$, so that our choice of form for an amplitude prior at the
Earth does not translate back into a luminosity distribution that varies with
distance.

One difficulty with any power-law prior is that it diverges for $s \to 0$. In a
reasonable scenario of SNR $>$a few, the integral over $s$ will consist of two
distinct contributions. The first is a Gaussian component with $\mu =
s_\mathrm{best}$ and variance $\sigl^2$, scaled by the prior
$P(s_\mathrm{best})$.  The second component is an infinite contribution from
the divergent prior at $s = 0$ with little contribution afterward due to
suppression from the Gaussian tail.  The infinite contribution represents the
certainty of a signal of arbitrarily small amplitude to be present in the data,
regardless of the ability of the data to resolve it. For moderate SNR, it's
very easy to isolate only the Gaussian contribution by placing a small cut on
amplitude, truncating the likelihood for $s < s_\mathrm{cut}$. Under the
approximation that $P(s)$ varies slowly over the width of the Gaussian, the
log-likelihood marginalized over all $s$ becomes,
\begin{equation}
    \label{eqn:marginalizedloglikelihood}
    \mathcal{L}(d) = -\beta\ln{s_\mathrm{best}} + \ln{\sigl} + \mathcal{L}(d|s_\mathrm{best})
\end{equation}
up to common additive constants that do not depend on the data $d$.

\begin{figure}
    \begin{center}
    \includegraphics[width=1.55in]{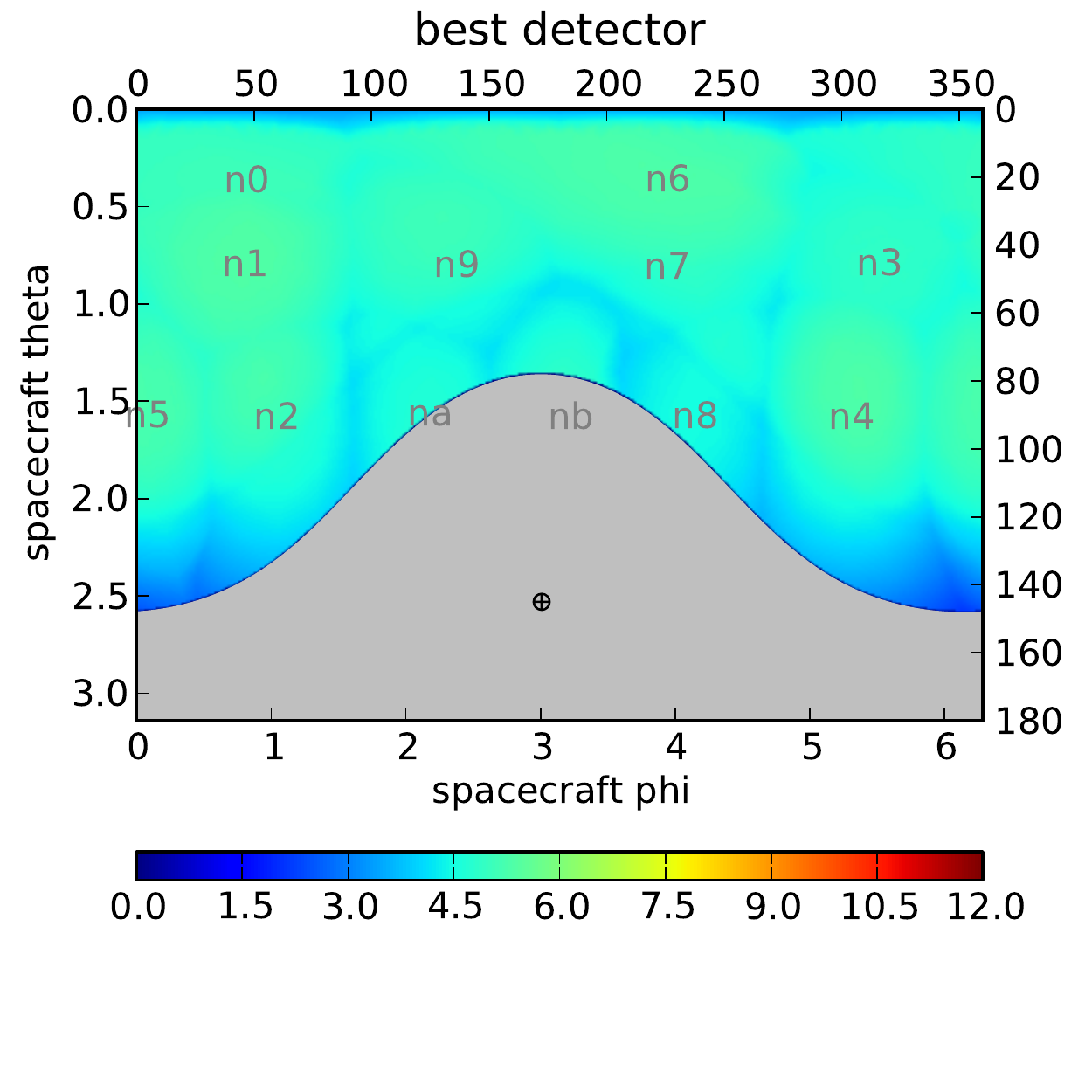}
    \includegraphics[width=1.55in]{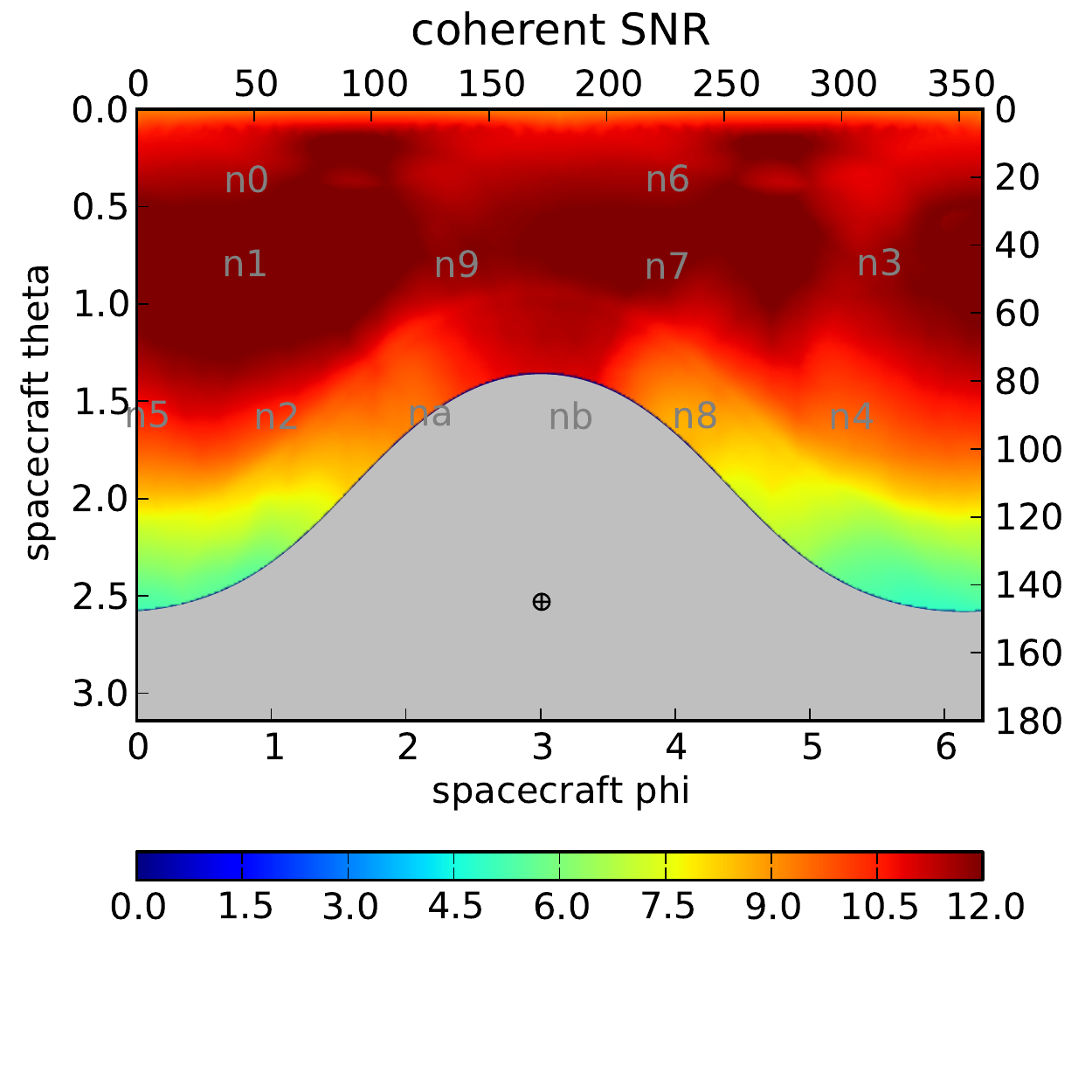}
    \caption{Signal-to-noise expected from GRBs with a normal spectrum, and
        coming from various positions across the GBM field-of-view. The
        hypothetical signal lasts 0.512s and is normalized to 1.0
        photons/s/cm$^2$ in the 50--300 keV band, while background rates and
        Earth position are selected from 10 seconds prior to GRB 090305A. The
        instrumental response includes contributions from atmospheric
        scattering in the 50--300 keV band. The first map shows the
        signal-to-noise from the single best detector over 50--300 keV. The
        second plot shows the SNR expected from a coherent analysis of the full
        CTIME data. The coherent statistic must be calculated for all
        directions, which means it is subject to a trials factor equal to the
        number of independent sky locations.}
    \label{fig:cohresp}
    \end{center}
\end{figure}

The approximation for the marginalized likelihood becomes problematic for small
$s_\mathrm{best}$ where the assumption that $P(s)$ is constant over the range
of the Gaussian breaks down, and the Gaussian distribution can no longer be
isolated from the divergence at small $s$. We enforce a finite and well-behaved
prior by multiplying by a prefactor,
\begin{equation}
    P(s) = \left[1-e^{-\left(s/\gamma\sigl\right)^\beta}\right] s^{-\beta}
\end{equation}
so that $P(s)$ reaches a maximum constant value of $(1/\gamma\sigl)^\beta$ for
small $s$. The tunable parameter $\gamma$ sets the number of standard
deviations at which the prior begins to plateau, and we use $\gamma = 2.5$.
This allows us to use the approximation that $P(s)$ is reasonably constant over
a range of $\sigl$ for any $s>0$. The only remaining correction is to account
for clipping of the Gaussian for non-physical $s<0$, which can be represented
by the error function. The final approximation for the amplitude-marginalized
log-likelihood becomes,
\begin{eqnarray}
    \nonumber
    \mathcal{L}(d) = \ln\sigl + 
    \ln\left[1+\mathrm{Erf}\left(\frac{s_\mathrm{best}}{\sqrt{2}\sigl}\right)\right]
    + \mathcal{L}(d|s_\mathrm{best}) \\
	+ \left\{
    \begin{array}{ll}
        \ln\left[1-e^{-\left(s_\mathrm{best}/\gamma\sigl\right)^\beta}\right] - \beta\ln{s_\mathrm{best}} & s_\mathrm{best} > 0 \\
        -\beta\ln\left(\gamma\sigl\right) & s_\mathrm{best} \leq 0
    \end{array}\right.
\end{eqnarray}
which contains factors form the Gaussian width, fractional overlap with $s >
0$, maximum likelihood at $s_\mathrm{best}$, and scaling from $P(s)$
respectively. Finally we are free to calibrate the log-likelihood by
subtracting the expected $\mathcal{L}(d)$ calculated for no signal at a
reference sensitivity: $\mathcal{L}_\mathrm{ref} = -\beta\ln\gamma +
(1-\beta)\ln\sigma_{\mathcal{L},\mathrm{ref}}$. $\sigl$ represents the source
amplitude required for a $1\sigma$ excess in the combined data, and is around
0.05 photons/s/cm$^2 \times (T/1$s$)^{-1/2}$ [50--300 keV] for typical source
spectra and reference background levels. Figure \ref{fig:cohresp} shows the
coherent signal-to-noise expected from all detectors for a 0.512s-long event
with normal GRB spectrum and constant amplitude of 1.0 photons/s/cm$^2$, and
compares it to the SNR expected from the most favorably-oriented individual
detector alone in the 50--300 keV band.

\section{Performing the follow-up}
\label{sec:followup}

For a single gravitational-wave trigger, we search a standard sGRB accretion
timescale of [0,~5s] relative to the time of coalescence for prompt flux excess
in GBM between 0.256 and 2s long (the lower limit of 0.256s set by the CTIME
archival resolution does not apply to continuous TTE data). We also search a
prior interval [--30s,~0s] for possible precursor bursts between 0.256s and 2s.
Finally we may include bursts between 2s to 32s in an extended interval
[--30s,~300s] to search for possible longer-duration emission. While emission
outside of the standard accretion timescale can be considered speculative, it's
worth looking for given that any events detectable in gravitational-waves will
be closer than sGRB's with known red-shift to date, making them good candidates
to search for weak exotic and possibly less-beamed emission. To appropriately
tile the search in time and duration $T$, we use rectangular windows with $T$
spaced by powers of two (0.256s, 0.512s, 1s, etc.).  Their central times are
sampled along the search interval in units of $T/4$ to provide an even mismatch
in signal-to-noise across search windows.

The likelihood-ratio statistic described in the previous section is calculated
for each foreground interval based on the background-subtracted flux
measurements $\tilde{d_i}$ in each of the 8 $\times$ 14 channel and detector
combinations. In addition, the instrumental response $r_i$ (and thus the
likelihood-ratio) also depend on source sky location and spectrum. To
efficiently calculate the response over a large area of sky, we use precomputed
all-sky response look-up tables originally generated for offline localization
\citep{Paciesas:2012vs}. The likelihood ratio as a function of sky position
provides a probability distribution over the sky of a GBM signal (as in figure
\ref{fig:weakevent}). This can be coincided with the GW-derived skymap (figure
\ref{fig:localization}) by direct multiplication, in effect using the GW skymap
as a prior.

Finally, we marginalize over sky location and representative source spectra to
get a ranking of events characterized only by their foreground time intervals.
A unique list of non-overlapping events is constructed by beginning with the
highest-ranked event and removing from the list any lower-ranked events with
foreground windows that overlap it; then taking the next surviving highest
ranked event, etc., until the list is exhausted.

\section{Test on Swift short GRBs}
\label{sec:test}

\begin{figure}[t]
\begin{center}
    \includegraphics[width=2.75in]{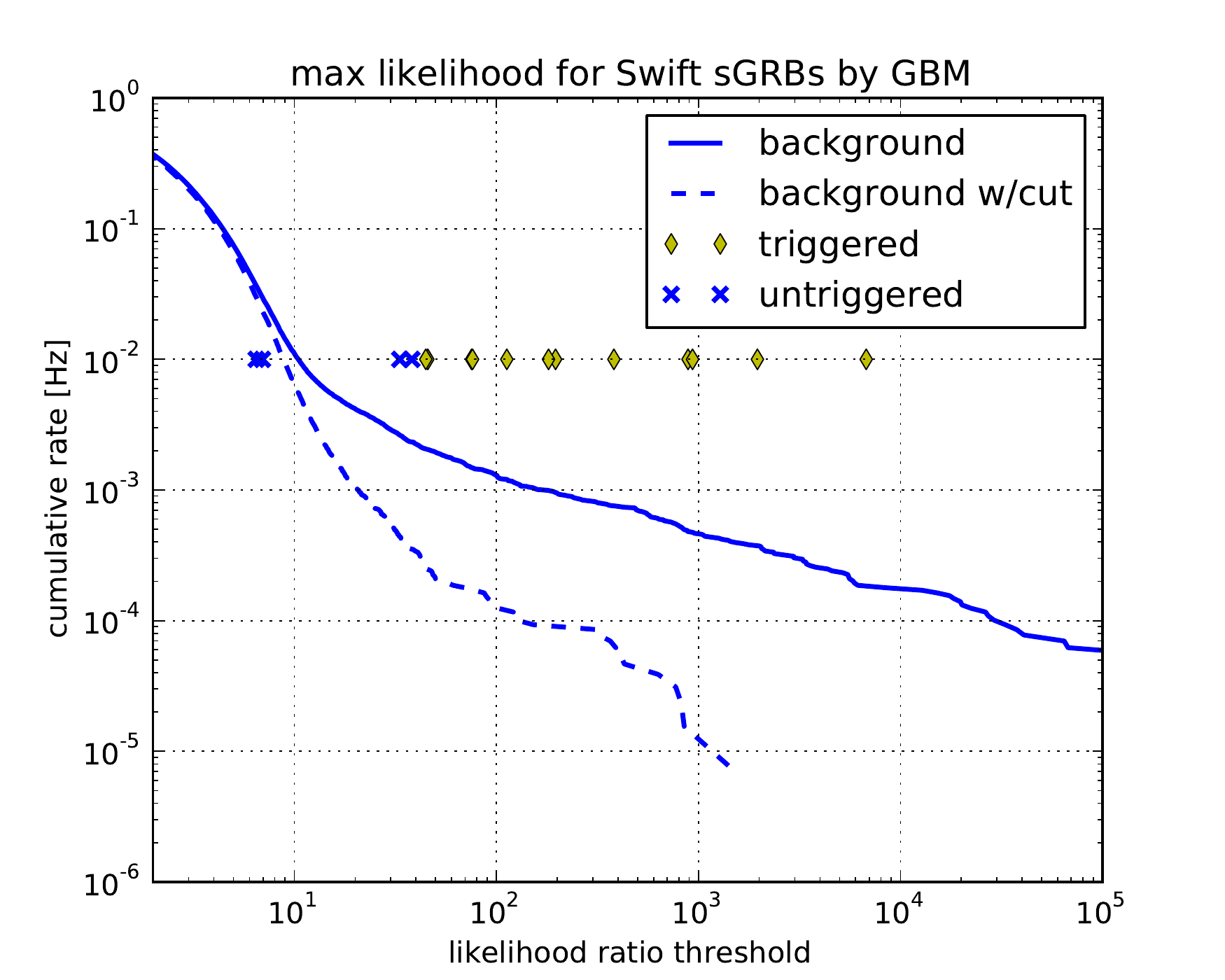} \caption{The maximum
        likelihood coherent flux excess found by searching time around Swift
        short-GRBs within the GBM field of view using foreground intervals
        between 0.256 and 2s. GRBs are placed at 0.01 Hz in anticipation of a
        GW-GBM coincidence search window of $<$100s. An estimated cumulative
        background rate distribution is found by taking the non-overlapping
        all-sky events from a continuous scan of approximately one-day in total
        of nearby data. The dotted background distribution shows the effect of
        a simple cut to remove particle events (the cut does not remove any
        GRBs). Of the four un-triggered events, GBM excesses corresponding to
        GRB 090815C and 110112A are buried within the background distribution,
        and are not clearly associable with their GRBs. The ability of the
        method to identify the next two weak, un-triggered events GRBs 090305A
        and 120403A (see also figure \ref{fig:weakevent}) depends on ongoing
        strategies to further reject non-Gaussian outliers. The remaining Swift
        sGRB's in the sample have a corresponding GBM on-board trigger.}
    \label{fig:bgdistribution}
\end{center}
\end{figure}

We can test the offline analysis on short GRBs triggered by Swift and
observable by Fermi-GBM. The Swift GRBs are particularly useful as their
accurate localizations resolve systematic errors in the GBM model response. For
this test, we calculate the all-sky marginalized likelihood ratio for GBM data
local to Swift sGRB measurements (figure \ref{fig:bgdistribution}). Here we
assume no previous information about source location. A background rate
distribution is obtained by running the same search over nearby time.

\begin{figure}[thb]
\begin{center}
    \includegraphics[width=2.75in]{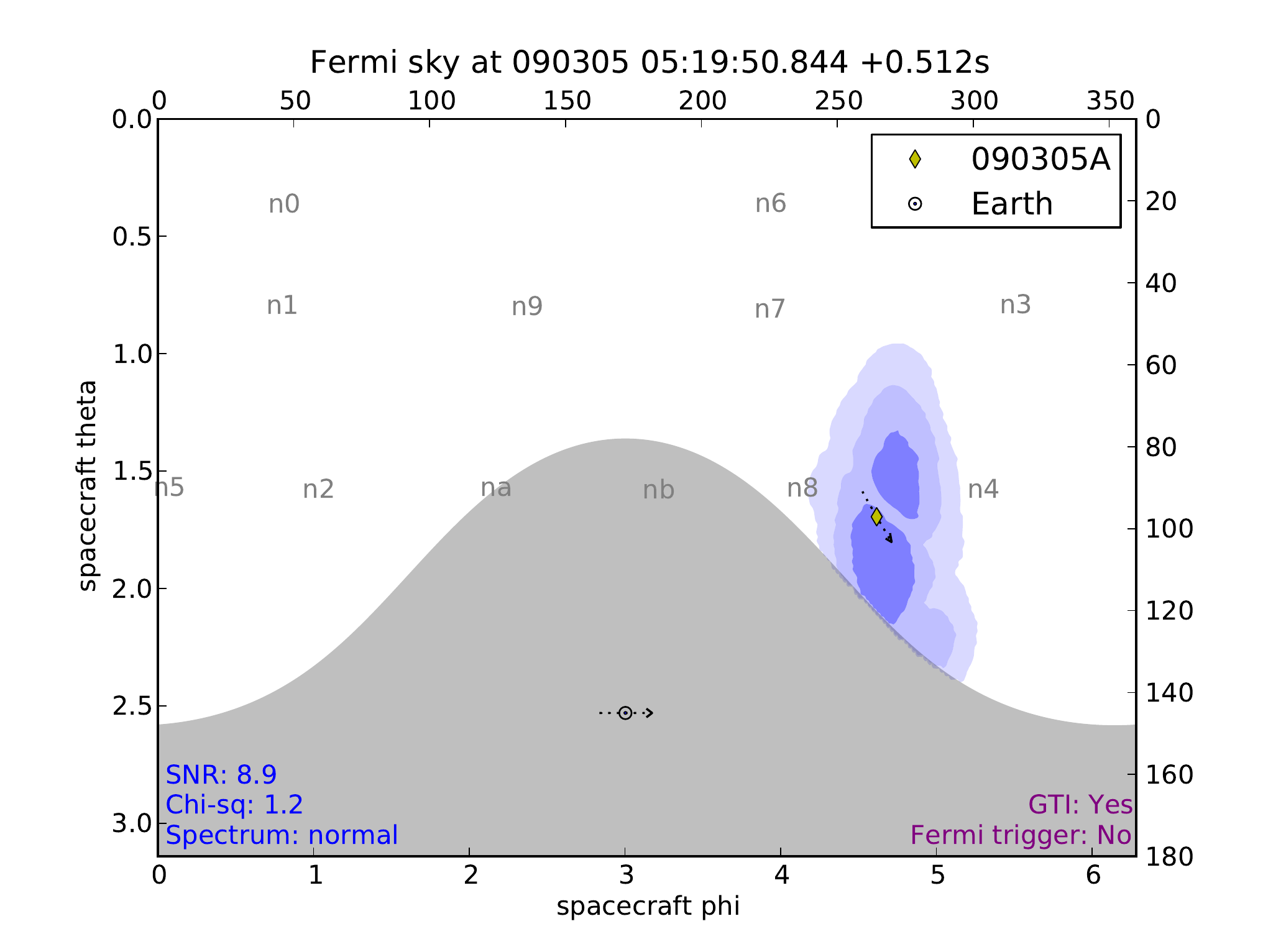}
    \includegraphics[width=2.75in]{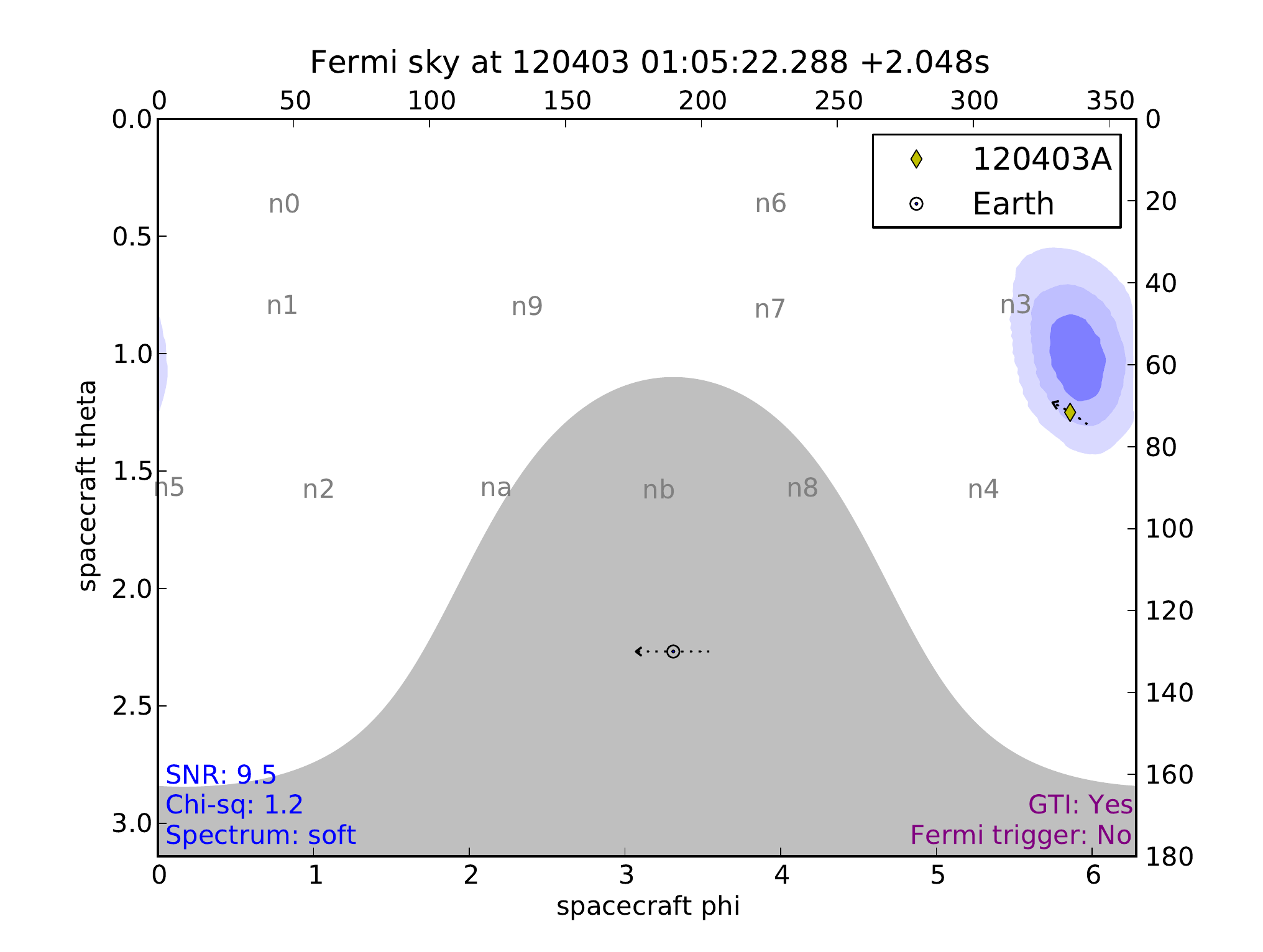} \caption{Swift
        short GRBs (T90 $<$ 2s) 090305A and 120403A were seen by Swift-BAT but
        did not trigger GBM on-board. Low photon statistics for these weak
        events result in large sky location uncertainties shown by the 1, 2,
        and 3$\sigma$ confidence regions. The quoted signal-to-noise ratio is
        calculated at the best-fit location using the full spectral data from
        all NaI and BGO detectors. The spectrum represents one of three
        representative spectral models (hard, normal, soft) for the model
        response that gives maximum likelihood, while the skymap is
        marginalized over all spectral models.  Arrows represent movement of
        the Fermi spacecraft over $\pm$2 minutes.}
    \label{fig:weakevent}
    \end{center}
\end{figure}

About 50\% of Swift sGRB's (T90~$<$~2s) are within GBM's field-of-view (the
65\% of the sky not occulted by the Earth) and occur during an interval of time
when Fermi is outside of the South-Atlantic Anomaly (SAA) and operational. Most
of these observable Swift sGRB's also triggered GBM on-board. GRB 081024A
triggered on-board, but was too close to the interruption of data taking during
passage through the SAA for our two-sided background estimation. GRB 090305A
and 120403A did not trigger on-board, and do not show compelling evidence for a
signal in the offline data. GRB 090305A and 120403A also did not trigger
on-board, but do show clear evidence of a signal present in the offline
analysis (figure \ref{fig:weakevent}), though with relatively poor statistics.
The remaining Swift short-GRBs from the observable sample both triggered
on-board and are clearly identified offline.

\section{Conclusion}

Direct detection of gravitational-waves from the merger of NS-NS or NS-BH
binary systems is expected to occur within the next few years as advanced 2nd
generation ground-based gravitational-wave detectors come online. Fermi GBM
provides a unique and promising opportunity to observe EM counterparts due to
its large sky coverage, and the anticipated association between NS-NS mergers
and sGRB's. To maximize the use of information from both GWs and GBM, we have
outlined a joint search strategy triggered by observations of GW signals from
coalescing binary systems in which a small amount of local GBM data is scanned
using a likelihood-ratio based analysis applied to the full instrument data.
The sensitivity of the method to short GRBs will benefit in the future from
continuous GBM offline data products with higher time resolution and improved
all-sky response models.  We anticipate such an effort will allow sensitive
follow-up of NS-NS and NS-BH mergers, most of which are expected to not be
accompanied by triggered GRB observation.

\section{Acknowledgments}

We thank Colleen Wilson-Hodge for assistance with the GBM direct response model.
LB is supported by an appointment to the NASA Postdoctoral Program at GSFC,
administered by Oak Ridge Associated Universities through a contract with NASA.
The authors would also like to acknowledge the support of the NSF through grant
PHY-1204371. Finally we thank the Albert Einstein Institute in Hannover,
supported by the Max-Planck-Gesellschaft, for use of the Atlas high-performance
computing cluster.

\bibstyle{apsrev}
\bibliography{symposium}{}

\end{document}